# Identification of active slip systems in polycrystals by Slip Trace - Modified Lattice Rotation Analysis (ST-MLRA)

Biaobiao Yang [a, b, c], Chenying Shi [a], Ruilin Lai [a], Dongfeng Shi [b, c], Dikai Guan [d], Gaoming Zhu [e], Yujie Cui [f, *], Guoqiang Xie [g], Yunping Li [a, *], Akihiko Chiba [f], Javier LLorca [b, c, *]

[a] State Key Lab for Powder Metallurgy, Central South University, Changsha 410083, China
[b] IMDEA Materials Institute, C/Eric Kandel 2, Getafe, 28906 - Madrid, Spain
[c] Department of Materials Science, Polytechnic University of Madrid/Universidad Politécnica de Madrid, E.T.S. de Ingenieros de Caminos, 28040 - Madrid, Spain
[d] Department of Materials Science and Engineering, University of Sheffield, Mappin Street, Sheffield S1 3JD, United Kingdom
[e] National Engineering Research Center of Light Alloy Net Forming, Shanghai Jiao Tong University, Shanghai 200240, China
[f] Institute for Materials Research, Tohoku University, Sendai 980-8577, Japan
[g] School of Materials Science and Engineering, Harbin Institute of Technology (Shenzhen), Shenzhen 518055, China
* Corresponding authors. E-mail: cuiyujie@imr.tohoku.ac.jp (Y. Cui); lyping@csu.edu.cn (Y. Li); javier.llorca@upm.es (J. LLorca).

**Abstract**: A simple and versatile strategy, denominated Slip Trace - Modified Lattice Rotation Analysis (ST-MLRA), is presented to enable the identification of the active slip systems in polycrystalline alloys from surface information. The slip plane trace orientation is used to define the potential active slip planes while the actual slip direction within the active slip plane is obtained from the grain rotation, as indicated by the stretching of the trace of the grain orientation in the pole figure as a result of deformation. Examples of application of the strategy in a HCP Mg alloy are presented for illustration. They show that the strategy is simple to implement and allows to identify the active slip system(s) in each grain.

Dislocation slip is the most important plastic deformation mechanism in metallic alloys and it has been widely investigated in the past [1–3]. In particular, the precise identification of the active slip system(s) is critical to understand the deformation mechanisms (slip localization, forest hardening, interaction of dislocations with grain boundaries, etc.) that arise during plastic deformation. So far, slip trace analysis in combination with electron backscatter diffraction (EBSD) information is the most common approach to identify the active slip system(s). Its kernel is the comparison of the observed slip trace on the crystal surface (i.e. the intersection between slip plane and crystal surface) with the predicted one taking into account the crystal orientation provided by EBSD [4]. This approach is enough when there is only one possible slip direction (slip system) in a given slip plane, e.g. $\{112\}$ and $\{12\overline{3}\}$ of body-centered cubic (BCC) and $\{1\overline{1}00\}$ and $\{11\overline{2}2\}$ of hexagonal close packed (HCP) lattices (Supplementary material Fig. 1). However, the active slip system cannot be identified with this strategy when there are 2 or more possible slip directions (slip systems) in a slip plane because all slip directions in a slip plane generate identical slip traces [5]. Such limitation of slip trace analysis is usually present in close-packed planes such as $\{111\}$ of face-centered cubic (FCC), $\{110\}$ of BCC and $\{0001\}$ of HCP lattices. Under these circumstances, it is necessary to rely on other criteria or constraints to identify the actual active slip system within the slip plane. They include, for instance, the maximum Schmid factor ($m$), the minimum critical resolved shear stress (CRSS), or the highest geometric compatibility factor ($m'$) between slip systems at a grain boundary [5–8]. Unfortunately, these criteria are potential indicators rather than sufficient conditions to identify the active slip system and may lead to erroneous conclusions [9,10].

Recently, a modified lattice rotation analysis -that enables precise identification of the rotation axis (RA) for the active slip system- was proposed [11]. This analysis relies on the EBSD images after deformation and is based on the comparison between the experimental and calculated evolution of the grain orientation due to deformation, as indicated by the projection in the pole figure. However, one rotation axis can be shared by 2 or 3 slip systems and the active slip system cannot be discriminated by the modified lattice rotation analysis. This is the case, for instance, of <a> basal slip in the {0001} plane and <c + a> pyramidal II slip in {11$\bar{2}$2}, that share the same <1$\bar{1}$00> rotation axis family in the HCP lattice, leading to identical evolution of the grain orientation due to plastic deformation. Other methodologies, such as *in situ* high resolution digital image correlation (HRDIC), 3D X-ray diffraction and transmission electron microscopy, can also be used to identify the active slip system during plastic deformation [12–16]. For example, the active slip system can be identified with the aid of HRDIC by coupling the information provided by the slip traces with information about the displacement fields around the slip traces obtained from HRDIC [12]. However, these techniques are time-consuming and require sophisticated procedures, which significantly restrict their widespread application.

In this investigation, a new and efficient strategy, denominated Slip Trace - Modified Lattice Rotation Analysis (ST-MLRA), is presented to identify the current active slip system in each grain during plastic deformation of polycrystalline metallic alloys. The input information are the Bunge Euler angles (**g** and **g'** matrices) from the region near the observed slip traces before and after deformation, respectively, and the orientation of the slip trace (given by the slip trace vector $\vec{T}$) in the sample surface. The first step in the analysis is to determine the slip plane. The vector normal to any slip plane ($\vec{SP}_{sc}$) in the sample coordinate (sc) system can be calculated as [17]:

$$\vec{SP}_{sc} = \vec{SP}_{cc} \cdot \mathbf{g} \tag{1}$$

where $\vec{SP}_{cc}$ stands for the vector normal to the slip plane in the crystal coordinate (cc) system. Note that the scalar and cross products of $\vec{A}$ and $\vec{B}$ are expressed by $\vec{A} \cdot \vec{B}$ and $\vec{A} \times \vec{B}$, respectively. The slip trace vector ($\vec{T}$) on the sample surface can be computed from the cross product of $\vec{SP}_{sc}$ and the vector normal to sample surface ($\vec{N}$)

$$\vec{T} = \vec{SP}_{sc} \times \vec{N} \tag{2}$$

and the comparison of $\vec{T}$, computed from eq. (2), with the experimental $\vec{T}$ determined from the scanning electron microscope (SEM) images indicates the actual slip plane.

The second step is to determine the slip direction within the slip plane. To this end, the Bunge Euler angles of the crystal after rotation (**g′** matrix) can be determined as the scalar product of the **g** matrix and rotation matrix **rot** as

$$\mathbf{g′} = \mathbf{g}\cdot\mathbf{rot} \qquad (3)$$

where **rot** matrix is calculated with MATLAB [TM] using the function of rotation.byAxisAngle(axis, angle) in MTEX [18] that takes into account the rotation axis vector $\overrightarrow{RA}_{sc}$ in the sample coordinate system and the rotation angle $\theta$ according

$$\mathbf{rot} = \text{rotation.byAxisAngle}(\overrightarrow{RA}_{sc}, \theta). \qquad (4)$$

. The rotation axis vector depends on the slip plane and slip direction [19] and can be computed as

$$\overrightarrow{RA}_{sc} = \overrightarrow{RA}_{cc}\cdot\mathbf{g} \qquad (5)$$

$$\overrightarrow{RA}_{cc} = \overrightarrow{SP}_{cc}\times\overrightarrow{SD}_{cc} \qquad (6)$$

where $\overrightarrow{RA}_{cc}$ and $\overrightarrow{SD}_{cc}$ stands for the rotation axis and slip direction vectors in crystal coordinate (cc) system respectively, as summarized in the Supplementary material Table 1. The initial orientation of the crystal in the pole figure before deformation (given by **g**) and after deformation (given by **g'**) should be connected by a line that is given by eq. (3) and that indicates which one is the actual slip direction within the slip plane. The rotation angle $\theta$ is normally $\leq 5°$ based on slip-induced local lattice rotations [11,20].

The feasibility of the ST-MLRA strategy to identify the actual slip system is demonstrated below for a HCP Mg alloy, a good example because of the large number of slip systems. The Mg-0.3Zn (at.%) alloy ingot was prepared by vacuum melting, followed by a homogenization treatment at 400°C for 2 h and furnace cooling. They were extruded at 300˚C with an extrusion ratio of 16:1 and a ram speed of $\approx$ 2 mm·s⁻¹, followed by another homogenization heat treatment at 400˚C for 2 h and furnace cooling. A dog-bone-shaped tensile sample with the gauge dimensions of $10 \times 2 \times 2$ mm³ (length × width × thickness) was manufactured by electron discharge machining with the tensile axis (TD) parallel to extrusion direction (ED). The sample surface was mechanically polished using SiC paper, $0.5, 0.05$ μm diamond slurries, and 40 nm oxide suspensions. Slight etching using 5% Nital solution (a mixture of nitric acid and ethanol) was then conducted for a few seconds on the sample surface. The tensile tests were performed using Shimadzu screw-driven tensile testing machine with a displacement speed of 0.05 mm·min⁻¹ at room temperature. The strain was measured using a non-contact digital video extensometer. The sample surface was analyzed in a SEM using secondary electrons (FEI XL30S, FEI Company, Portland, OR, USA; spot size: 3.5, accelerating voltage: 20 kV) and by EBSD (Oxford HKL Channel 5, Oxford Instruments, Abingdon, UK; step size: 0.5 μm). During ST-MLRA analysis, it was assumed that plastic slip in HCP Mg could take place along 16 slip planes (numbered from 1 to 16) and the corresponding rotation axes for all the slip directions in all slip planes (denominated from A to V). This detailed information can be found in Supplementary material Table 1.

The application of ST-MLRA to identify the active slip system in grain G1 with Euler angles of (165.4°, 64.7°, 48.8°) is depicted in Fig. 1. The observed slip traces in G1 (Fig. 1a) are in good agreement with the calculated one for the (0001) basal plane (number 1, Fig. 1b). However, there are 3 slip directions in this plane: $[2\bar{1}\bar{1}0]$ (Schmid factor $m$ = 0.20, Rotation axis (RA): $[01\bar{1}0]$ (A)), $[\bar{1}2\bar{1}0]$ ($m$ = 0.15, RA: $[10\bar{1}0]$ (B)), and $[\bar{1}\bar{1}20]$ ($m$ = 0.05, RA: $[1\bar{1}00]$ (C)), and the active one cannot be discriminated by conventional slip trace analysis. The Schmid factor ($m$) for slip system is calculated by:

$$m = \cos(\varphi) \cdot \cos(\lambda) \tag{7}$$

where $\varphi$ and $\lambda$ stand for the angle between slip plane normal direction with the slip direction and the TD, respectively. Obviously, the Schmid factor is calculated neglecting local effects associated with the deformation of neighboring grains that may change the actual stress state in the grain. These effects are particularly significant at large applied strains (10%) but can also be important at smaller strains and, thus, the Schmid factor -as defined in eq. (7)- is not always an accurate criterion to predict the most active slip system within a slip plane.

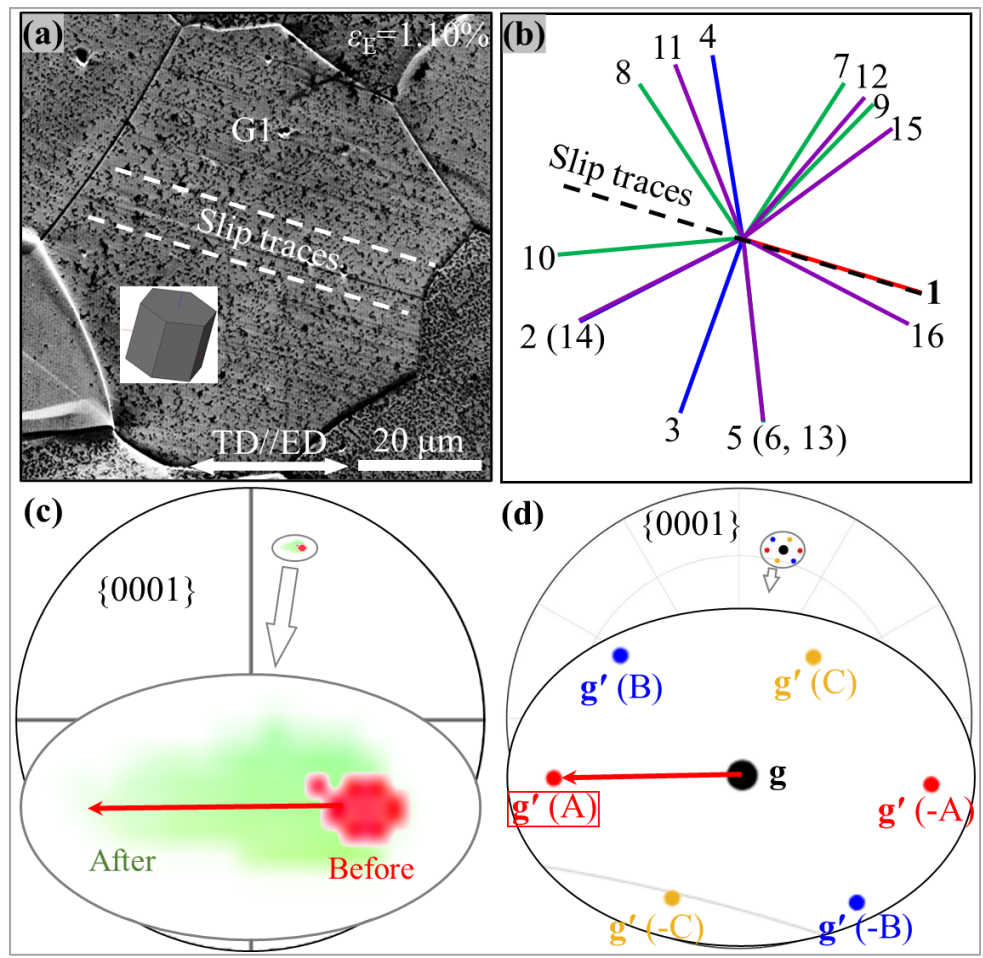

Figure 1. (a) Experimental slip traces observed by SEM on grain 1 (G1) of the Mg alloy surface (applied tensile strain 1.10%). (b) Calculated slip traces of different slip planes according to the EBSD information from G1. (c) Experimental {0001} pole figure of grain G1 before (red) and after (green) deformation. The main stretching direction of the trace of the grain is shown with a red arrow. (d) Simulated projections of grain G1before (**g**) and after (**g′**) a rotation of 5º around various possible rotation axes. The projection points of the **g′** matrix are shown in different colors corresponding to different rotation axes: red (RA: ± [01$\bar{1}$0] / A), blue (RA: ± [10$\bar{1}$0] / B), yellow (RA: ± [1$\bar{1}$00] / C).

The experimental projection of the orientation of grain G1 in the pole figure before and after deformation is plotted in Fig. 1c. Plastic deformation stretches the projection in the horizontal direction which is in good agreement with the predictions corresponding to the rotation axis $[01\bar{1}0]$ (A) and far away from those predicted for rotation axes $[10\bar{1}0]$ (B) or $[1\bar{1}00]$ (C), as shown in Fig. 1d. Therefore, it can be concluded that the only active slip system in G1 is the basal <a> (0001) $[2\bar{1}\bar{1}0]$ which, in this case, is consistent with the result predicted by the maximum Schmid factor criterion.

A more complex scenario is found in grain G2 with Euler angles (167.6°, 90.4°, 13.7°) (Fig. 2). As shown in Figs. 2a and b, the experimental slip traces are compatible with two possible active slip planes: $(0\bar{1}11)$ pyramidal I (number 10) and $(1\bar{2}1\bar{2})$ pyramidal II (number 13). There are 3 possible slip directions in the $(0\bar{1}11)$ pyramidal I plane: $[2\bar{1}\bar{1}0]$ (<a> dislocation, $m = 0.13$) with RA $[01\bar{1}2]$ (T), $[11\bar{2}3]$ (<c + a> dislocation, $m = 0.08$) with RA $[3\bar{2}\bar{1}\bar{1}]$ (U), and $[1\bar{2}1\bar{3}]$ (<c + a> dislocation, $m = 0.01$) with RA $[3\bar{1}\bar{2}1]$ (V). However, there is one slip system in the $(1\bar{2}1\bar{2})$ pyramidal II plane: $(1\bar{2}1\bar{2})$ $[1\bar{2}13]$ (<c + a> dislocation, $m = 0.02$) with RA $[10\bar{1}0]$ (B). These four possible slip systems cannot be discriminated using conventional slip trace analysis. In fact, the normal procedure to determine the active slip plane using standard slip trace analysis is to adopt an arbitrary "tolerance" angle to discard the potential slip planes that are not within this tolerance [12,13]. However, the deviation between the experimental and the predicted slip traces for both slip planes are very similar in this case: $\alpha = 10.7°$ for $(0\bar{1}11)$ pyramidal I (number 10) and $\beta = 10.8°$ for $(1\bar{2}1\bar{2})$ pyramidal II (number 13) (Fig. 2b).

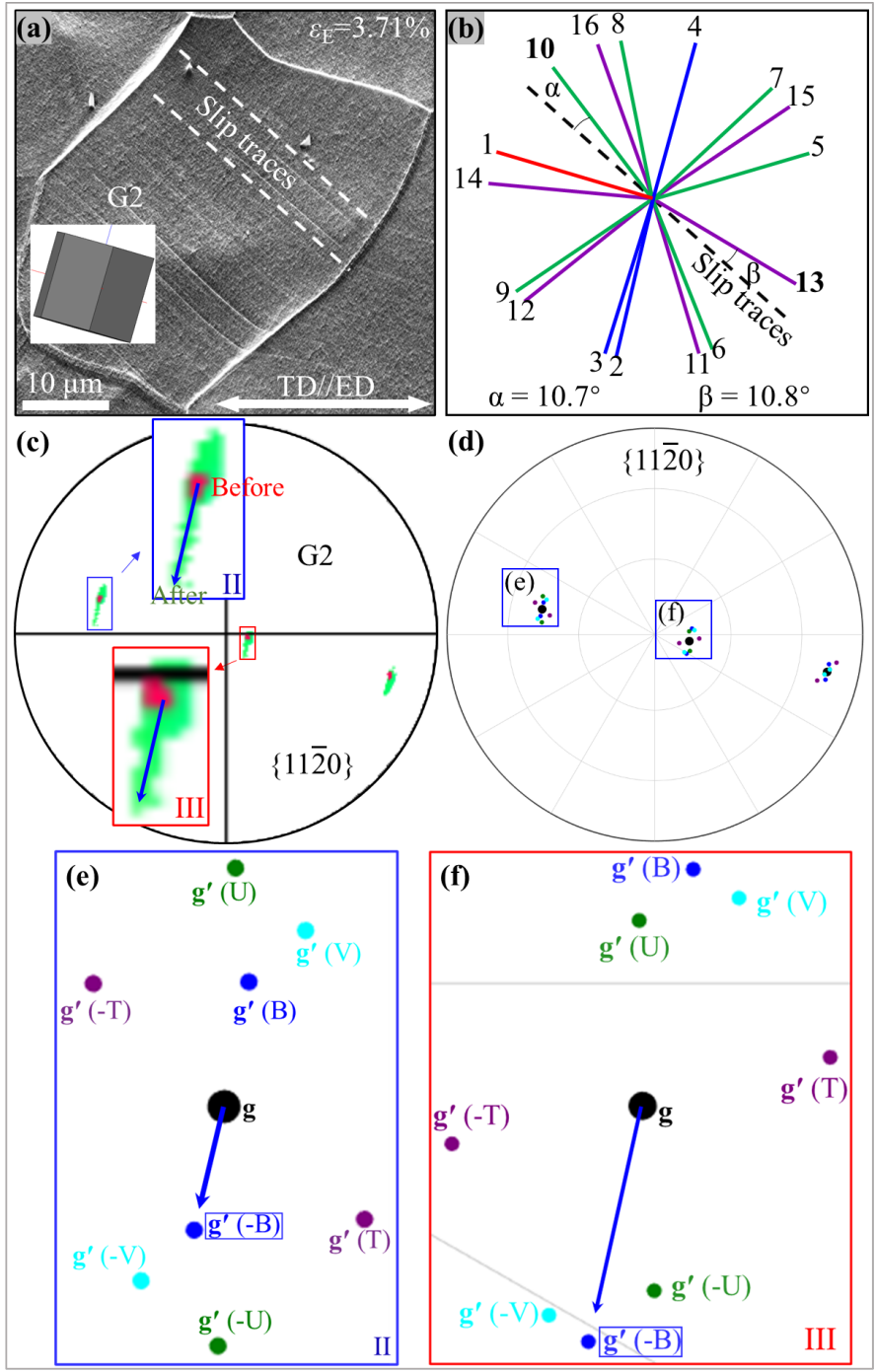

Figure 2. (a) Experimental slip traces observed by SEM in G2 of the Mg alloy surface (applied tensile strain 3.7%). (b) Calculated slip traces of different slip planes according to the EBSD information from G2. (c) Experimental $\{11\bar{2}0\}$ pole figure of grain G2 before (red) and after (green) deformation. Note that the $\{11\bar{2}0\}$ rather than the $\{0001\}$ pole figure was selected because the projection of grain G2 in the $\{0001\}$ pole figure was too close to the edge. The main stretching directions of the traces of the grain are shown with blue arrows. (d) Simulated projections of grain G2 before (**g**) and after (**g′**) 5º rotation around various axes based on the Euler angles. Enlarged views of regions I and II in Fig. 2d are shown in (e) and (f), respectively. The projection points of the **g′** matrix are shown in different colors corresponding to the different rotation axes: blue (RA: $\pm [10\bar{1}0]$ / B), purple (RA: $\pm [01\bar{1}2]$ / T), green (RA: $\pm [3\bar{2}\bar{1}\bar{1}]$

/ U), cyan (RA: $\pm\,[3\overline{1}\overline{2}1]$ / V).

The experimental projection of the orientation of grain G2 in the pole figure before and after deformation is plotted in Fig. 2c while the predictions for the stretching of the projections corresponding to the four possible rotation axes are plotted in Figs. 2d, e and f. It is evident that the best match between experiments and simulations is found for the RA $[10\overline{1}0]$ (B) and, thus, plastic deformation in grain G2 took place along the <c + a> $(1\overline{2}1\overline{2})$ $[1\overline{2}13]$ pyramidal slip system. The large deviation angle between the orientation of the experimental and calculated slip trace (10.85°) in Fig. 2b should be related to the rotation of grain during the tensile tests [21]. Thus, ST-MLRA allows to find out the correct slip system when the slip traces are slightly misoriented with respect to the theoretical ones due to strain gradients associated with the polycrystalline deformation.

Planar slip in a single slip system was dominant in the examples in Figs. 1 and 2 but double or multiple slip is also a possibility that should be accounted for. For instance, the experimental slip traces in grain G3 (Euler angles = (170.9°, 73°, 0.9°)) are not parallel to one another (Fig. 3a). They are compatible with three possible active slip planes: (0001) basal (number 1), $(1\overline{2}1\overline{2})$ pyramidal II (number 13) and $(\overline{1}2\overline{1}2)$ pyramidal II (number 14) (Fig. 3b). There are 3 possible slip directions in the (0001) basal plane: $[2\overline{1}\overline{1}0]$ ($m = 0.13$ with RA $[01\overline{1}0]$ (A)), $[\overline{1}2\overline{1}0]$ ($m = 0.01$ with RA $[10\overline{1}0]$ (B)), and $[\overline{1}\overline{1}20]$ ($m = 0.13$, with RA $[1\overline{1}00]$ (C)). However, there is only one slip direction in the $(1\overline{2}1\overline{2})$ pyramidal II and $(\overline{1}2\overline{1}2)$ pyramidal II planes: $[1\overline{2}13]$ (<c + a>) ($m = 0.01$, with RA $[10\overline{1}0]$ (B)) and $[1\overline{2}1\overline{3}]$ (<c + a>) ($m = 0.01$, with RA $[10\overline{1}0]$ (B)), respectively. The active slip systems cannot be discriminated among these 5 possibilities using standard slip trace analysis.

The experimental projection of the orientation of grain G3 in the pole figure before and after deformation is plotted in Fig. 3c while the predictions for the stretching of the projections corresponding to the three possible rotation axes are plotted in Fig. 3d. The best match between the experimental stretching of projections and the simulations is found for the rotation axis $[1\bar{1}00]$ (C) and, thus the main active slip system in G3 is basal <a> (0001) $[\bar{1}\bar{1}20]$ slip system. However, the experimental projection of grain G3 was also stretched in two other directions, which were compatible with the activation of the other two basal <a> slip systems (with RA $[01\bar{1}0]$ (A) or RA $[10\bar{1}0]$ (B)), and/or <c + a> $(1\bar{2}1\bar{2})$ $[1\bar{2}13]$ and $(\bar{1}\bar{2}12)$ $[1\bar{2}1\bar{3}]$ pyramidal II slip systems with RA $[10\bar{1}0]$ (B). The intensity of the slip activity in each slip system can be estimated by the rotation angle necessary to move the initial orientation of grain G3 to the final one in the experimental pole figure in each of the three orientations. They lead to 9.2° for the rotation around $[1\bar{1}00]$ (C), 3.8° around $[01\bar{1}0]$ (A) and 5.7° along $[10\bar{1}0]$ (B), which provides an estimation of the relative activity of each slip system.

The precise identification of slip system is very important, for instance, in order to assess the geometrical factors that determine whether slip transfer or blocking occurs at grain boundaries. This information can be used within the framework of crystal plasticity to develop microstructure-based models for polycrystalline deformation [22,23]. Among the different criteria to assess slip transfer at grain boundaries, the geometric compatibility factor or Luster-Morris parameter ($m'$) is one of the most relevant ones [24–28]. It is based upon the angles between the slip plane normal directions $\psi$ and the Burgers vector directions $\kappa$ according to

$$m' = (\overrightarrow{SP_{\text{in}}} \cdot \overrightarrow{SP_{\text{out}}}) \cdot (\overrightarrow{SD_{\text{in}}} \cdot \overrightarrow{SD_{\text{out}}}) = \cos(\psi) \cdot \cos(\kappa) \tag{8}$$

where $\overrightarrow{SP_{\text{in}}}$, $\overrightarrow{SP_{\text{out}}}$, $\overrightarrow{SD_{\text{in}}}$, and $\overrightarrow{SD_{\text{out}}}$ represents the slip planes normal vectors and slip directions of incoming and outgoing slip systems, respectively (Supplementary material Figure 2).

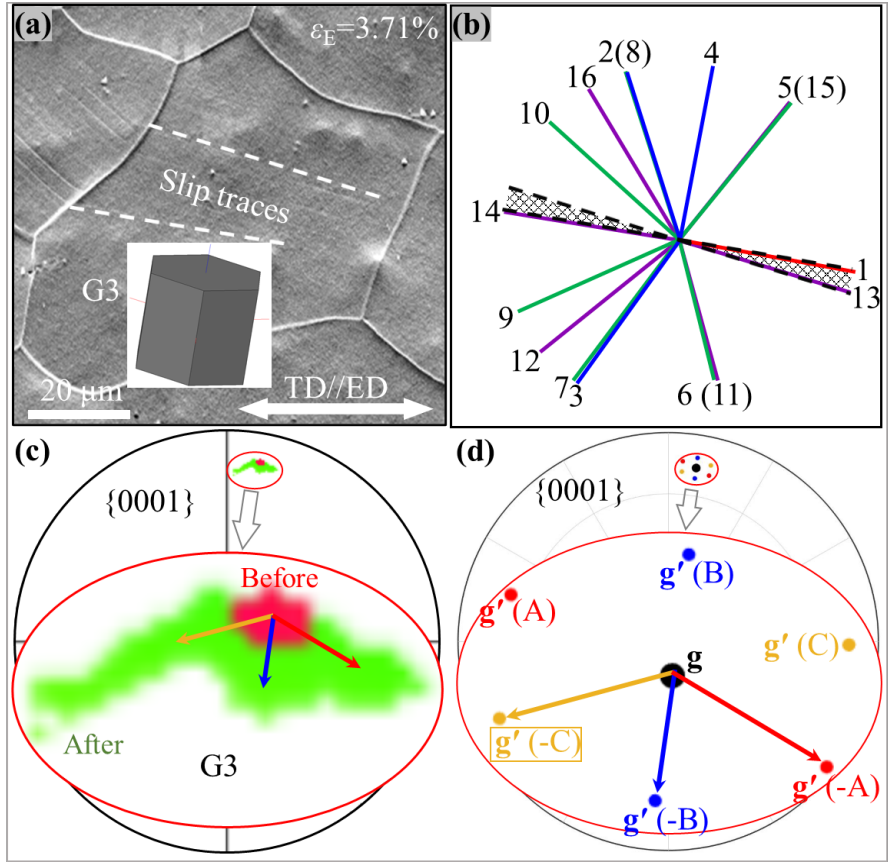

Figure 3. (a) Experimental slip traces observed by SEM in grain G3 of the Mg alloys surface (applied tensile strain 3.7%). (b) Calculated slip traces of different slip planes according to the EBSD information from G3. (c) Experimental {0001} pole figure of grain G3 before (red) and after (green) deformation. (d) Simulated projections of grain G3 before (**g**) and after (**g′**) 5º rotation around various axes based on the Euler angles. The projection points of the **g′** matrix are shown in different colors corresponding to the different rotation axes: red (RA: ± [01$\bar{1}$0] / A), blue (RA: ± [10$\bar{1}$0] / B), yellow (RA: ± [1$\bar{1}$00] / C).

The SEM micrograph in Fig. 4 shows slip traces in neighbor grains G2 and G3. The presence of a ledge at the boundary and the lack of correlation between most of the slip traces in both grains indicate that slip transfer did not take place and that both two grains deformed heterogeneously. The active slip system in grain G2 was identified as (1$\bar{2}$1$\bar{2}$) [1$\bar{2}$13] with a Schmid factor of 0.02 (Fig. 2). The main active slip system in G3 was identified as (0001) [$\bar{1}\bar{1}$20] with a Schmid factor of 0.13 (Fig. 3) and the *m′* parameter corresponding to both slip systems is 0.25 (highlighted in bold in Table 1). This low *m′* value is consistent with the lack of slip transfer at the grain boundary [17], as indicated by the SEM observations. It should be noted, however, that if all the possible active slip systems -according to standard slip trace analysis- are included in Table 1, high values of $m'$ ($> 0.80$) can be found for several pairs of slip systems ($m' = 0.887$ for $(1\bar{2}1\bar{2})$ $[1\bar{2}13]$ in grain G2 and $(1\bar{2}1\bar{2})$ $[1\bar{2}13]$ in grain G3 and $m' = 0.836$ for $(0\bar{1}11)$ $[1\bar{2}1\bar{3}]$ in grain G2 and $(\bar{1}2\bar{1}\bar{2})$ $[1\bar{2}1\bar{3}]$ in grain G3) (Table 1). Note these pyramidal slip systems are hard to be activated because of the large CRSS and, in addition, their Schmid factor is very low ($m < 0.02$). Thus, it is very likely that pyramidal slip in G2 was activated as a result of the stress concentration at the grain boundary induced by the pile-up of the basal dislocations in G3.

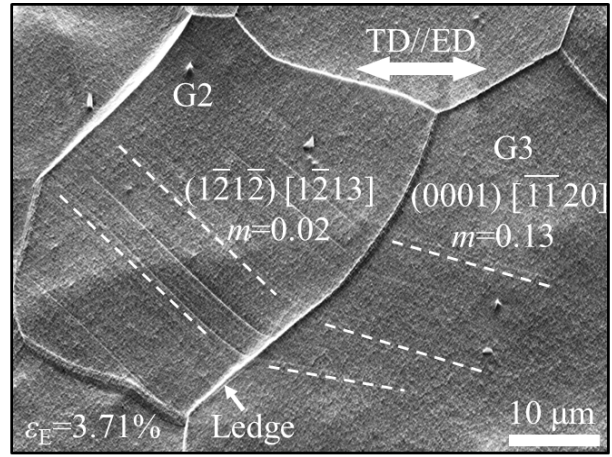

Figure 4. Example of slip system identification using ST-MLRA strategy for neighbor grains G2 and G3 in the Mg alloy to determine the geometric computability factor for slip transfer (applied tensile strain 3.7%). The main active slip systems are shown in each grain.

Table 1. Schmid factors, $m$, for slip systems in grains G2 (G2ss) and G3 (G3ss), and geometric compatibility factor $m'$ across the boundary for the different possible active slip systems identified by ST-MLRA. The highlighted slip systems are expected to be dominant in G2 and G3 according to ST-MLRA. The $m'$ value corresponding the main active slip systems in each grain is highlighted.

| $m'$ | G2ss | $(0\bar{1}11)$ $[2\bar{1}\bar{1}0]$ / 10T | $(0\bar{1}11)$ $[11\bar{2}3]$ / 10U | **$(1\bar{2}1\bar{2})$ $[1\bar{2}13]$ / 13B** | $(0\bar{1}11)$ $[1\bar{2}1\bar{3}]$ / 10V |
|---|---|---|---|---|---|
| G3ss | $m$ | 0.13 | 0.08 | **0.02** | 0.01 |
| $(0001)$ $[2\bar{1}\bar{1}0]$ / 1A | 0.13 | 0.177 | 0.054 | 0.404 | 0.076 |
| **$(0001)$ $[\bar{1}\bar{1}20]$ / 1C** | **0.13** | 0.123 | 0.102 | **0.250** | 0.012 |
| $(0001)$ $[\bar{1}2\bar{1}0]$ / 1B | 0.01 | 0.054 | 0.048 | 0.654 | 0.088 |
| $(1\bar{2}1\bar{2})$ $[1\bar{2}13]$ / 13B | 0.01 | 0.233 | 0.259 | 0.887 | 0.156 |
| $(\bar{1}2\bar{1}\bar{2})$ $[1\bar{2}1\bar{3}]$ / 14B | 0.01 | 0.216 | 0.677 | 0.056 | 0.836 |

Thus, a strategy to determine the active slip system in each grain during plastic deformation of polycrystalline has been presented. The strategy, denominated, Slip Trace-Modified Lattice Rotation Analysis (ST-MLRA), is based on the determination of the orientation of the slip traces on the grain surface in combination with the changes in the grain orientation given by the EBSD maps before and after deformation. The slip traces indicate the potential active slip planes while the stretching of the trace of the grain orientation in the pole figure confirms the actual rotation axes in the potential slip planes. The strategy was demonstrated in grains of Mg polycrystals subjected to relatively low tensile strains ($<$ 5%) which show single slip or multiple slip but with one dominant slip system. The accuracy may be reduced at high strains because more slip systems are active and more slip traces emerge to surface, leading to diffuse and/or wavy slip traces. Thus, identification of the actual active slip planes becomes more difficult and the combinations of the rotations of associated with the different slip systems make less obvious the stretching of the trace in the pole figure along well-defined directions.

The strategy can be very easily implemented and provides the precise identification of the active slip system(s) in many cases from simple surface observations with secondary electrons and EBSD in the scanning electron microscope. Thus, large data sets with the precise active slip system in each grain can be obtained, which will enhance our understanding the fundamental deformation mechanisms of polycrystals.

**Acknowledgements**: This investigation was supported by the grant from the Natural Science Foundation of China (51871244), the Hunan Provincial Innovation Foundation for Postgraduate (CX20200172) and the Fundamental Research Funds for the Central Universities of Central South University (1053320190103). Mr. B. Yang wishes to express his gratitude to Mr. Shaolou Wei from Massachusetts Institute of Technology for the assistance on slip trace analysis and the support by China Scholarship Council (202106370122). Dr. D. Guan would like to thank the MRC for his UKRI Future Leaders Fellowship, MR/T019123/1. Prof. J. LLorca acknowledges the support from the the HexaGB project of the Spanish Ministry of Science and Innovation (reference RTI2018-098245, MCIN/AEI/ 10.13039/501100011033).

## References

[1]  W. Li, D. Xie, D. Li, Y. Zhang, Y. Gao, P.K. Liaw, Progress in Materials Science 118 (2021) 100777.

[2]  Z. Wu, W.A. Curtin, Nature 526 (2015) 62–67.

[3]  B.-Y. Liu, F. Liu, N. Yang, X.-B. Zhai, L. Zhang, Y. Yang, B. Li, J. Li, E. Ma, J.-F. Nie, Z.-W. Shan, Science 365 (2019) 73–75.

[4]  C.M. Cepeda-Jiménez, J.M. Molina-Aldareguia, M.T. Pérez-Prado, JOM 68 (2016) 116–126.

[5]  Y. Guo, D.M. Collins, E. Tarleton, F. Hofmann, A.J. Wilkinson, T.B. Britton, Acta Materialia 182 (2020) 172–183.

[6]  Z. Huang, L. Wang, B. Zhou, T. Fischer, S. Yi, X. Zeng, Scripta Materialia 143 (2018) 44–48.

[7]  D.D. Yin, C.J. Boehlert, L.J. Long, G.H. Huang, H. Zhou, J. Zheng, Q.D. Wang, International Journal of Plasticity 136 (2021) 102878.

[8]  A. Githens, S. Ganesan, Z. Chen, J. Allison, V. Sundararaghavan, S. Daly, Acta Materialia 186 (2020) 77–94.

[9]  B. Šesták, N. Zárubová, Physica Status Solidi (b) 10 (1965) 239–250.

[10] R. Gröger, V. Vitek, Materials Science Forum 482 (2005) 123–126.

[11] B. Yang, C. Shi, X. Ye, J. Teng, R. Lai, Y. Cui, D. Guan, H. Cui, Y. Li, A. Chiba, Journal of Magnesium and Alloys (2021).

[12] X. Xu, D. Lunt, R. Thomas, R.P. Babu, A. Harte, M. Atkinson, J.Q. da Fonseca, M. Preuss, Acta Materialia 175 (2019) 376–393.

[13] D. Lunt, R. Thomas, M.D. Atkinson, A. Smith, R. Sandala, J.Q. da Fonseca, M. Preuss, Acta Materialia 216 (2021) 117111.

[14] B. Zhou, L. Wang, W. Liu, X. Zeng, Y. Li, Metall Mater Trans A 51 (2020) 4414–4421.

[15] C. Marichal, H. Van Swygenhoven, S. Van Petegem, C. Borca, Sci Rep 3 (2013) 2547.

[16] G. Zhu, L. Wang, H. Zhou, J. Wang, Y. Shen, P. Tu, H. Zhu, W. Liu, P. Jin, X. Zeng, International Journal of Plasticity 120 (2019) 164–179.

[17] T.R. Bieler, R. Alizadeh, M. Peña-Ortega, J. Llorca, International Journal of Plasticity 118 (2019) 269–290.

[18] F. Bachmann, R. Hielscher, H. Schaeben, Solid State Phenomena 160 (2010) 63–68.

[19] D.E. Laughlin, K. Hono, Physical Metallurgy, 5th ed, Newnes, 2014.

[20] B. Zhou, L. Wang, P. Jin, H. Jia, H.J. Roven, X. Zeng, Y. Li, International Journal of Plasticity 128 (2020) 102669.

[21] Z. Chen, S.H. Daly, Exp Mech 57 (2017) 115–127.

[22] S. Haouala, R. Alizadeh, T.R. Bieler, J. Segurado, J. LLorca, International Journal of Plasticity 126 (2020) 102600.

[23] E. Nieto-Valeiras, S. Haouala, J. LLorca, European Journal of Mechanics - A/Solids 91 (2022) 104427.

[24] E. Bayerschen, A.T. McBride, B.D. Reddy, T. Böhlke, J Mater Sci 51 (2016) 2243–2258.

[25] B. Yang, C. Shi, S. Zhang, J. Hu, J. Teng, Y. Cui, Y. Li, A. Chiba, Journal of Magnesium and Alloys (2021).

[26] J. Luster, M.A. Morris, Metall Mater Trans A 26 (1995) 1745–1756.

[27] R. Alizadeh, M. Peña-Ortega, T.R. Bieler, J. LLorca, Scripta Materialia 178 (2020) 408–412.

[28] M.T. Andani, A. Lakshmanan, V. Sundararaghavan, J. Allison, A. Misra, Acta Materialia 200 (2020) 148–161.